\documentclass{article}

\usepackage{arxiv}

\usepackage[utf8]{inputenc} 
\usepackage[T1]{fontenc}    
\usepackage{hyperref}       
\usepackage{url}            
\usepackage{booktabs}       
\usepackage{amsmath,amsfonts,amssymb}       
\usepackage{nicefrac}       
\usepackage{microtype}      
\usepackage{graphicx}
\usepackage[backend=biber, style=numeric, sorting=none, citestyle=numeric]{biblatex}
\usepackage{doi}

\title{Wide Field of View Large Aperture Meta-Doublet Eyepiece}

\author{Anna Wirth-Singh$^{1,*}$\\
\And
Johannes E. Fr\"{o}ch$^{1,2}$\\
\And 
Fan Yang$^{3}$\\
\And
Louis Martin$^{3}$\\
\And
Hualiang Zhang$^{4}$\\
\And 
Quentin T. Tanguy$^{2}$\\
\And
Zhihao Zhou$^{2}$\\
\And
Luocheng Huang$^{2}$\\
\And
Demis D. John$^{5}$\\
\And
Biljana Stamenic$^{5}$\\
\And
Juejun Hu$^{3}$\\
\And
Tian Gu$^{3}$\\ 
\And
Arka Majumdar$^{1,2,**}$\\
\And
\\
$^{1}${Department of Physics, University of Washington, Seattle, WA 98195, USA}\\
$^{2}${Department of Electrical and Computer Engineering, University of Washington, Seattle, WA 98195, USA}\\
$^{3}${Department of Materials Science and Engineering, Massachusetts Institute of Technology, Cambridge, MA 02139}\\
$^{4}${Department of Electrical and Computer Engineering, University of Massachusetts, Lowell, MA 01854}\\
$^{5}${Department of Electrical and Computer Engineering, University of California, Santa Barbara, CA 93106}\\
$^{*}$\textit{annaw77@uw.edu}\\
$^{**}$\textit{arka@uw.edu}
}

\begin{document}
\maketitle

\begin{abstract}

Wide field of view and light weight optics are critical for advanced eyewear, with applications in augmented/virtual reality and night vision. Conventional refractive lenses are often stacked to correct aberrations at wide field of view, leading to limited performance and increased size and weight. In particular, simultaneously achieving wide field of view and large aperture for light collection is desirable but challenging to realize in a compact form-factor. Here, we demonstrate a wide field of view (greater than 60$^\circ$) meta-optic doublet eyepiece with an entrance aperture of 2.1 cm. At the design wavelength of 633 nm, the meta-optic doublet achieves comparable performance to a refractive lens-based eyepiece system. This meta-doublet eyepiece illustrates the potential for meta-optics to play an important role in the development of high-quality monochrome near-eye display and night vision systems.

\end{abstract}

\section{Introduction}

Alongside advances in artificial intelligence and widespread availability of digital content, demand for augmented reality (AR) and virtual reality (VR) near-eye displays has surged. There is great commercial interest in developing such technologies for education, gaming, and social interactions, and also significant defense and national security interest for night vision and enhanced vision. The human eye is a highly optimized system, so exceptional optical performance is required to facilitate the interaction between the user and virtual reality. For reference, human vision's full field of view is approximately $120 ^\circ$ \cite{Kress20}, which exceeds the performance of most "wide" (generally > $60 ^\circ$) angle camera systems. However, achieving such performance in near-eye displays presents significant optical engineering challenges \cite{Xiong21,Chang20}. In addition, near-eye optics must be thin and lightweight for user comfort and safety, especially for long-duration usage to reduce fatigue. With traditional optics, there is often a trade-off between compact form factor and performance, and the ultimate challenge of near-eye displays is that it demands both. 

A comfortable reading distance is around 35 cm \cite{Shieh07}, which is much greater than the distance between a head-mounted near-eye display and the eye. The challenge for near-eye optics, then, is to project the image on a display that is placed close to the eye to a comfortable viewing distance to avoid visual fatigue and discomfort \cite{Peli95,Chang20,Choi24}. Similarly, a common configuration for night vision goggles is to collect reflected moon and starlight in the near-infrared via an objective lens, intensify and upconvert that light to visible illumination via intensifier tubes, and couple that light into the eye via eyepiece optics \cite{Waxman98}. In both applications, eyepiece optics are required to collimate near-eye illumination in order to project the image to a comfortable viewing distance. It is desirable to mount such optics near the head to minimize torque on the wearer and maintain a compact form factor; on the other hand, the minimum acceptable physical distance between an optic and the surface of the eye (called eye relief) is about 1.5 cm. This provides a small amount of working space, thus requiring compact optics. "Pancake" lenses are commonly used, but suffer from low efficiency due to polarization conversions \cite{Cakmakci21}. Several emerging optical technologies are poised to meet the demands of near-eye displays, including holographic optical elements \cite{Chang20,Gao17,Gopakumar24} and leaky waveguides \cite{Jang24,Huang23}. Integrating metasurfaces directly onto micro-LED displays for collimation has also been proposed \cite{Chen24} but not yet experimentally demonstrated.

Ultra-thin and versatile meta-optics are another promising platform for near-eye displays. Meta-optics consist of arrays of sub-wavelength scatterers which impart spatially-varying phase shifts to incident light. Thanks to advancements in nanoscale lithography techniques, the fabrication of meta-optics at near-infrared and visible wavelengths is now regularly accomplished, with recent works demonstrating visible and near-infrared meta-optics with wide field of view \cite{Mart20,Shal20,Arba16,Yang23}, broadband performance \cite{Huang22,Froch24,Tseng21}, and large aperture \cite{Zhang23} for various applications, including AR/VR \cite{Li21,Li22Inverse,Lee18,Chen20,Bayati21}. In particular, the ability to achieve wide field of view in compact form-factor renders meta-optics particularly suitable for near-eye display applications. In singlet meta-optic lenses, nearly diffraction-limited performance has been experimentally demonstrated with over 170$^\circ$ field of view at mid-infrared and near-infrared wavelengths \cite{Shal20}, albeit with a relatively small entrance aperture of 1 mm ($\approx 200\lambda$). In another work, by relaxing the constraint on diffraction-limited performance, Martins et al. \cite{Mart20} demonstrate 178$^\circ$ field of view with larger entrance aperture of 2 mm ($\approx 3750\lambda$). 

\begin{figure}[h!]
\centering\includegraphics[width=12cm]{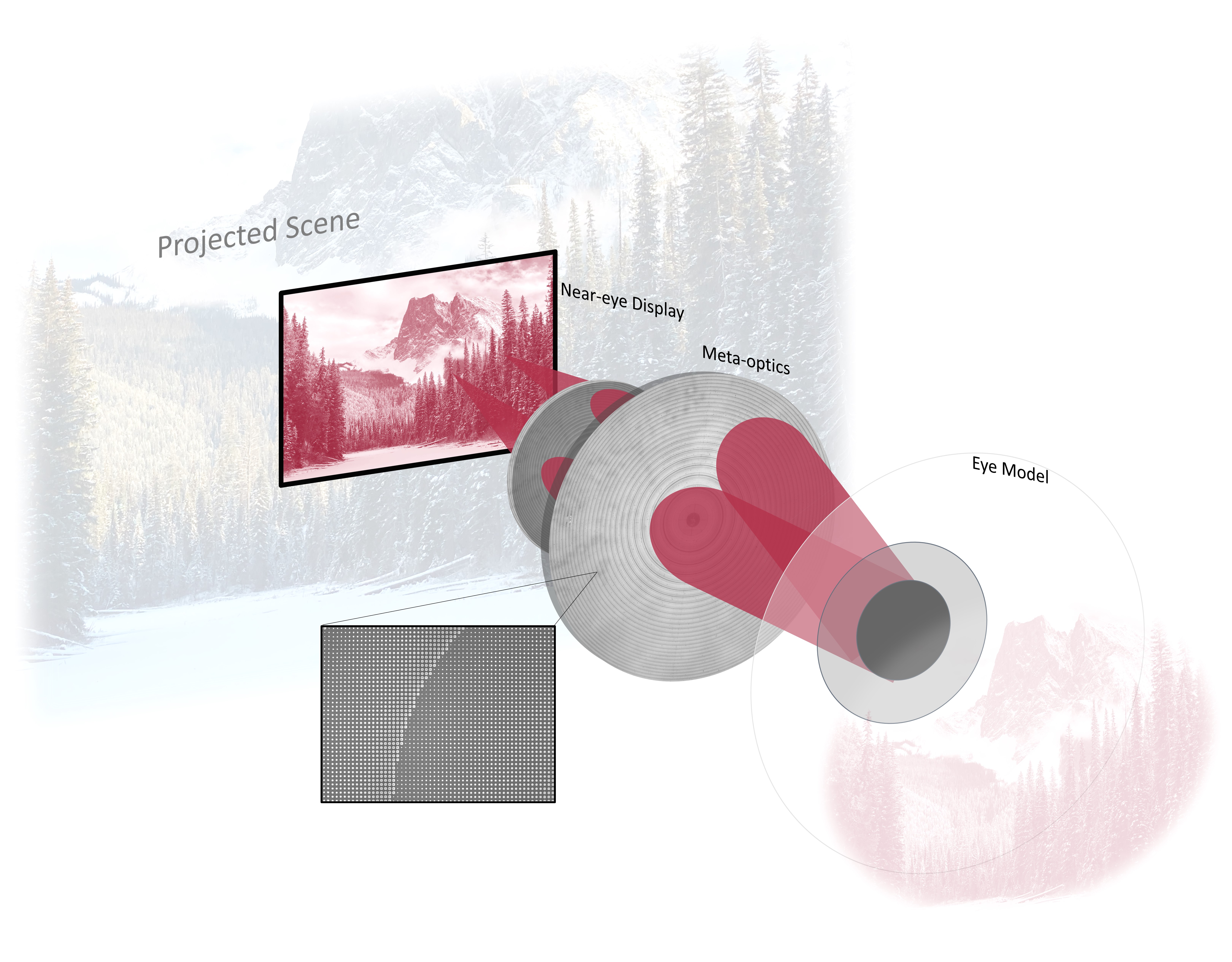}
\caption{Eyepiece schematic. Illumination from the near-eye display is collimated via two meta-optics into the pupil of the eye. The inset illustrates that the metasurface is composed of arrays of nanoscale pillars. }
\label{Fig:Schematic}
\end{figure}

 Similar to traditional refractive optics, stacking several lenses in doublet or triplet configurations provides more degrees of freedom for improved performance and additional functionalities. For example, meta-optic doublets have been used to correct monochromatic aberrations like spherical aberration, coma, and astigmatism as early as 2016 in near-infrared \cite{Arba16} and visible \cite{Groever17} wavelengths by patterning two metasurfaces on two sides of a single transparent substrate. While impressive, both of these designs had entrance apertures less than 1 mm, which is insufficient to be used as an eyepiece for near-eye displays. Since monochromatic Seidel aberrations scale with aperture size and field angle, achieving high optical quality is inherently difficult for simultaneously large aperture and wide field of view optics \cite{Lohm89}. In order to achieve diffraction-limited resolution over large field of view and relatively large aperture, it has been shown that doublet configurations are required \cite{Martins22}. In such a configuration, the first metasurface functions as both entrance aperture and a corrective plate, and the second functions a focusing lens.  

In this work, we demonstrate high quality image projection over $60^\circ$ field of view at 633 nm via a meta-optic doublet for eyepiece applications. We take a step-by-step approach towards realizing large aperture (2 cm, $\approx 31500 \lambda$) optics by first demonstrating the concept on a 1 cm aperture meta-optic doublet with $80^\circ$ field of view. We then demonstrate the full-scale system with large aperture (2 cm) and $60^\circ$ field of view. The system is designed for realistic eye relief (15 mm), pupil size (5.4 mm), and display size (16 mm). We show excellent experimental agreement with the theoretical model for both sets of optics. Further, compared to a similar commercially available refractive lens eyepiece system, the meta-optic system is superior in terms of improved image quality over wide field of view at the design wavelength and reduced total track length. This work closes the gap between previous wide field of view metalens demonstrations and practical applications. 

\begin{figure}[h!]
\centering\includegraphics[width=14cm]{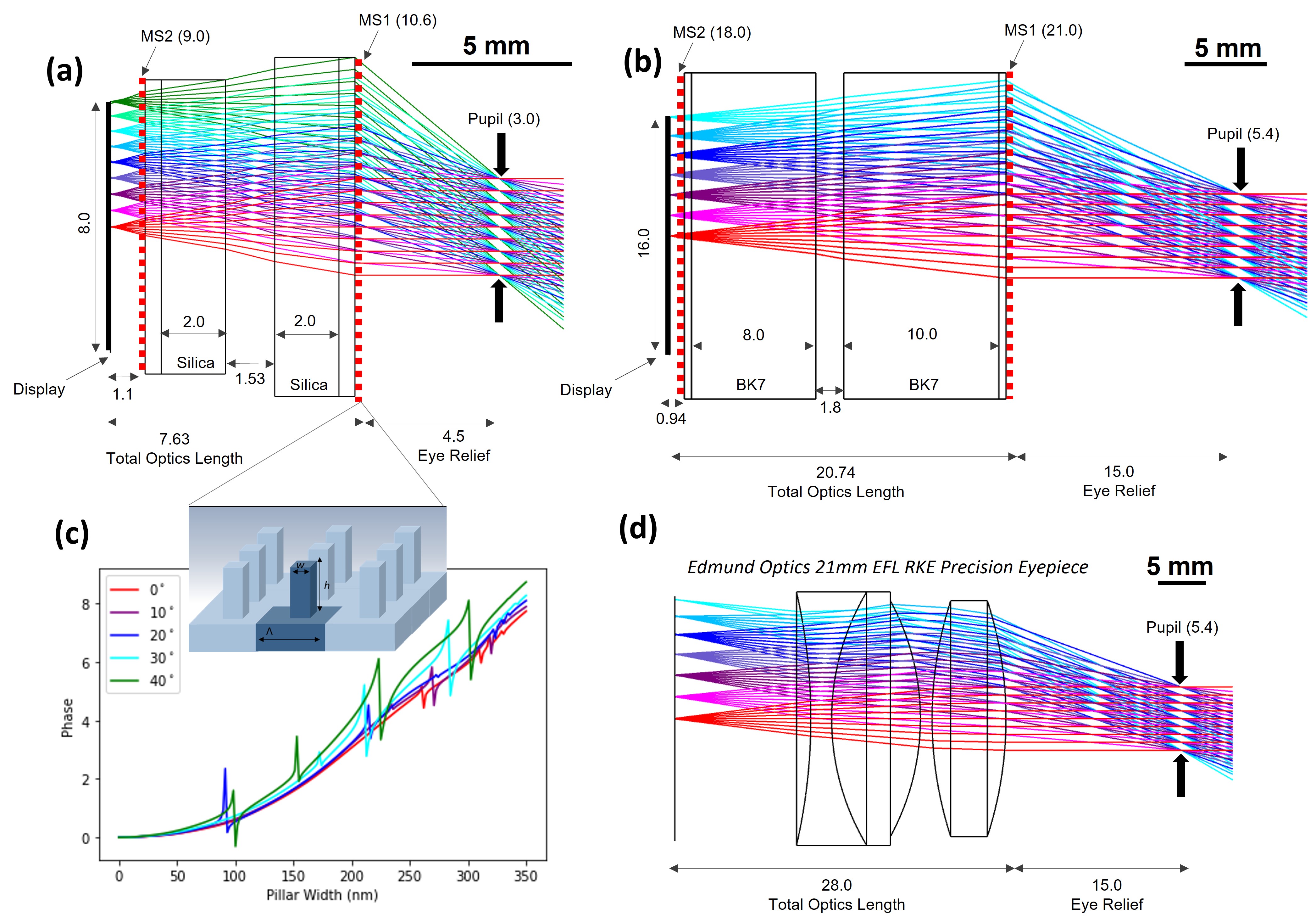}
\caption{Eyepiece ray tracing diagrams. All units are in mm unless otherwise specified. (a) Ray tracing diagram for the 1 cm aperture meta-optic doublet eyepiece with $80 ^\circ$ field of view. Rays are colored by field angle, with normally incident rays in red and rays at $40 ^\circ$ incidence in green. (b) Ray tracing diagram for the 2 cm aperture meta-optic doublet eyepiece with $60 ^\circ$ field of view. (c) Meta-atom design. The meta-optic consists of a periodic array (fixed periodicity $\Lambda$) of square pillars of uniform height $h$ and variable width $w$. Meta-atom phase responses are calculated using rigorous coupled wave analysis for various angles of incidence. (d) Ray tracing diagram for a comparable commercial refractive eyepiece. Angles of incidence up to $60 ^\circ$ are shown. The scale bar in (a), (b), and (d) is 5 mm.}
\label{Fig:RayDiagrams}
\end{figure}

\section{Results}

\subsection{Design and Modeling}

The concept of the wide field of view eyepiece is schematically illustrated in Figure \ref{Fig:Schematic}. The meta-optic is designed to collimate light from a near-eye display into the pupil with greater than 60$^\circ$ field of view. When the optics exhibit wide field of view functionality similar to that of the human eye, the user perceives a wide scene comfortably located at infinity, despite the display being physically located near the eye. 

In greater detail, ray tracing diagrams of the designed meta-optic doublet systems are shown in Figure \ref{Fig:RayDiagrams}. The 1 cm aperture design, shown in Fig. \ref{Fig:RayDiagrams}a, is demonstrated as a proof of concept prior to the full aperture (2 cm) design shown in Fig. \ref{Fig:RayDiagrams}b. In both cases, the system consists of a display (at the left), the two meta-optics (labeled MS1 and MS2) mounted on glass spacers with a small air gap in between, and the pupil aperture. Since beams with wide field of view diverge quickly, the required size of MS1 to collect the diverging beams rapidly increases; to mitigate this, optical windows (fused silica (n = 1.46) in the 1 cm designs and BK7 glass (n = 1.52) in the 2 cm designs) are used to reduce the beam divergence. Similar to other metasurface doublet systems \cite{Arba16,Groever17}, the entrance meta-optic (labeled MS2) functions as both an aperture stop and corrector plate, and the second meta-optic (labeled MS1) possesses the majority of the focusing power. The phase profiles were optimized using Zemax OpticStudio (more details in the Supplementary Information). We designed the system for 80 $^\circ$ full field of view in the 1 cm system and 60 $^\circ$ full field of view in the 2 cm system; in this case, the field of view is practically limited by the size of the second meta-optic due to fabrication constraints. 

\begin{table}[h]
\caption{Eyepiece Design Specifications}\label{tab:specifications}
\begin{tabular*}{\textwidth}{@{\extracolsep\fill}cccc} 
\toprule
  & 1 cm Meta-Optic Doublet & 2 cm Meta-Optic Doublet & Refractive Triplet \\
\midrule
Entrance Aperture (mm) & 10.6 & 21.0 & 20.0  \\
Designed field of view ($^\circ$) & 80 & 60 & 45* \\
Pupil Diameter (mm) & 3.0 & 5.4 & 5.4 \\
Eye Relief (mm) & 4.5 & 15.0 & 15.0 \\
Effective Focal Length (mm) & 5.84 & 15.17 & 21.61 \\
Numerical Aperture & 0.25 & 0.18 & 0.124 \\
Total Track Length (mm) & 12.1 & 35.7 & 43.0 \\
\bottomrule
\end{tabular*}
* Reported apparent field of view
\end{table}

As schematically illustrated in Figure \ref{Fig:RayDiagrams}c, the meta-optics consist of quasi-periodic arrays of rectangular pillars. To provide full $2\pi$ phase shift while maintaining high transmission, we use 750 nm tall silicon nitride (n = 2.04 \cite{Luke2015}) pillars with widths ranging from 80 to 270 nm on quartz (n = 1.46 \cite{Malitson1965}) substrate. The lattice periodicity is 350 nm and the expected transmission is greater than 80\% at normal incidence. To determine the corresponding phase delay of the structure, we used rigorous coupled wave analysis (RCWA) \cite{LiuV12} to calculate phase shift as a function of pillar width for various angles of incidence, as shown in Figure \ref{Fig:RayDiagrams}c. Due to the quasi-periodic nature, the phase shift from individual meta-atoms is not very sensitive to angle of incidence and thus we utilize the normally incident phase-width response to map between pillar width and phase when designing the optics. To reduce memory consumption during this mapping, we utilized the circular symmetry of the optic to generate a quarter of the radius width map and then copy and rotate the structure to produce the full circular optics. Some reduction in transmission is expected as the incident angle is increased; the calculated transmission for various angles of incidence is shown in Supplementary Figure S1b.

For comparison, we include a ray tracing diagram of a similar commercially available refractive lens eyepiece (Edmund Optics 66-210, 21 mm EFL RKE Precision Eyepiece) in Figure \ref{Fig:RayDiagrams}d. The refractive system has similar entrance aperture (20 mm) as our 2 cm meta-optic system and slightly longer effective focal length (21 mm for the refractive system versus 15 mm for the meta-optic system). We present both systems under the same pupil and eye relief conditions, namely 5.4 mm pupil diameter and 15 mm eye relief. From the ray tracing diagram, distortion is evident in the refractive system at wide field of view, whereas the designed meta-optic system exhibits very little distortion, even at incident angles of 30$^\circ$. In Table \ref{tab:specifications}, we summarize the key design specifications of the discussed systems, including their effective focal length, numerical aperture, total track length, and eye relief. The effective focal length and numerical aperture were calculated from the Zemax model. We show the simulated ray aberration and distortion curves in Supplementary Figure S2.

\subsection{Meta-optics Fabrication and Characterization}

\begin{figure}[h!]
\centering\includegraphics[width=12cm]{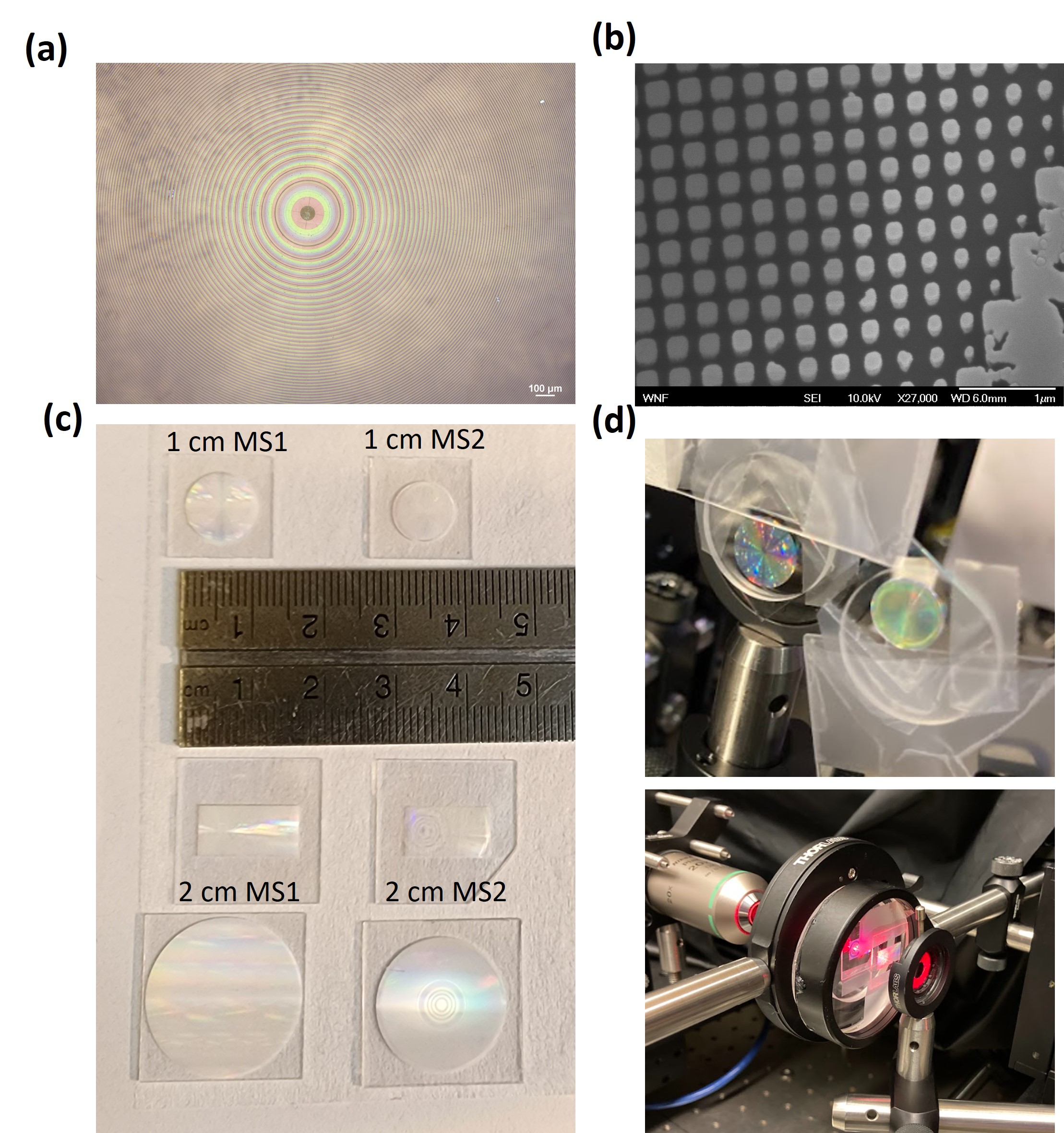}
\caption{Fabricated meta-optics. (a) Optical microscope image of the fabricated meta-optics. (b) SEM image of the fabricated optics at slightly oblique ($10^\circ$) viewing angle. (c) A photo of all fabricated meta-optics next to a ruler for scale. From top to bottom: the 1 cm aperture meta-optics, the 5.4 mm wide slices of the 2 cm meta-optics, and the full 2 cm aperture meta-optics fabricated using lower resolution scatterers. (d) A photograph of the 1 cm meta-optics (top) and 2 cm meta-optics (bottom) mounted on spacers in the experiment setup. }
\label{Fig:Fabrication}
\end{figure}

The meta-optics were fabricated in silicon nitride on quartz using electron beam lithography and inductively coupled plasma (ICP) fluorine etching, with further details in Methods. Optical microscope and scanning electron microscope (SEM) images of the fabricated devices are shown in Figures \ref{Fig:Fabrication}a and \ref{Fig:Fabrication}b to highlight the fabrication quality. Some of the larger pillars are not fully separated due to resolution constraints. The silica and glass spacers were obtained from commercial sources (2 mm thick fused silica: Newport FSW14; 8 mm thick BK7 glass: Thorlabs WG11508; 10 mm thick BK7 glass: Newport 20BW40-30). 

The fabricated optics, shown next to a ruler for scale, are shown in Figure \ref{Fig:Fabrication}c. The maximum aperture that we could fabricate using electron beam lithography was approximately 1 cm diameter, limited by the stability of the machine over extended write time. Therefore, the 2 cm eyepiece optics require a larger write area than realistically feasible using our methods. To circumvent this issue, we fabricated the entire aperture of the 1 cm meta-optics and only a slice of the 2 cm optics which was required to characterize the point spread function (PSF) of the optics. To measure the PSF, only the projected area of the pupil aperture (5.4 mm) is illuminated; therefore, the PSF characterization can be completed using only a slice of the metasurface with a dimension of 5.4 mm by 13.2 mm, covering the center to the outer edge. Therefore, we present PSF measurements of both the 1 cm and 2 cm optics and imaging results for the 1 cm optics only. In addition to electron beam lithography, we further describe fabrication of the full aperture 2 cm optics using deep ultraviolet (DUV) lithography; these optics are pictured at the bottom of Figure 3c and further details are provided in the Supplementary Information. While DUV lithography is a more scalable lithography process, the resolution of our process was limited to $\approx 250$ nm which is insufficient for sub-wavelength periodicity. The pictured full aperture 2 cm meta-optics are functional up to approximately $40^\circ$ full field of view, limited by aliasing issues arising from large periodicity (see the Supplementary Information for details).   

\begin{figure}[h!]
\centering\includegraphics[width=16cm]{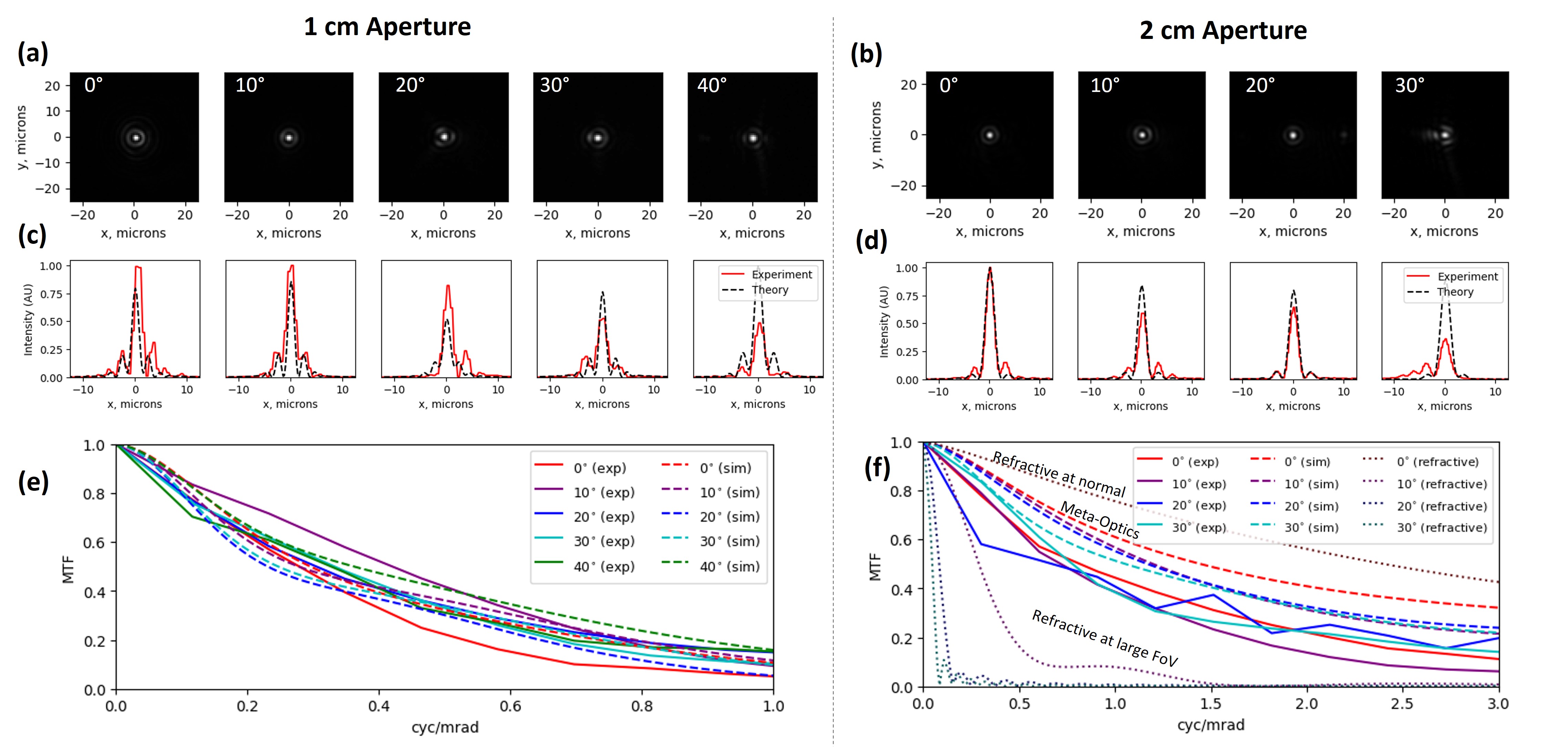}
\caption{Lens point spread function measurements. (a,b) Measured PSFs of (a) the 1 cm aperture optics from 0$^\circ$ to 40$^\circ$ degree angle of incidence, and (b) the 2 cm aperture optics from 0$^\circ$ to 30$^\circ$ degree angle of incidence. (c,d) Line cuts from the PSFs in (a) and (b) comparing the experimental results (solid red lines) to the theoretical results from the ray tracing model (dashed black lines). The camera exposure time was kept constant in each case. The simulated and experimental PSF results were normalized with respect to the maximum intensity value in the set. (e,f) The experimentally measured (solid lines) and simulated (dashed lines) MTF for (e) 1 cm meta-optics and (f) 2 cm meta-optics at increasing angles of incidence. In (f), the MTF of the comparable commercial refractive eyepiece is plotted as dotted lines.}
\label{Fig:PSFs}
\end{figure}

In Figure \ref{Fig:Fabrication}d, we show the fabricated optics in the experimental setup. To evaluate the performance of the fabricated meta-optics, we measured the PSF at various angles of incidence, using collimated output from a HeNe laser ($\lambda = 632.8$ nm, linewidth = 1400 MHz) as the illumination source. The simulated and measured PSFs are summarized in Figure \ref{Fig:PSFs}. In both the 1 cm and 2 cm aperture designs, the PSF (Figure \ref{Fig:PSFs}a and \ref{Fig:PSFs}b, respectively) remains mostly undistorted as the angle of incidence is increased. A horizontal line cut through each PSF is shown below, in Fig \ref{Fig:PSFs}c and \ref{Fig:PSFs}d, with the simulated PSF overlain in dashed black line, showing excellent agreement with the theory. For both the 1 cm and 2 cm systems, the PSF is normalized to the maximum measured intensity of that system. Some reduction in PSF intensity at larger angle of incidence is expected and observed, with the measured PSF intensity at 20 degrees off axis being 82\% and 64\% of that at normal incidence for the 1 cm and 2 cm designs, respectively. However, the PSF width remains mostly undistorted, highlighting the utility of the lens system over wide field of view. In addition, the measured transmission through the optics is consistent with that predicted by RCWA (see Supplementary Figure S1).  At normal incidence, the measured transmission through the 1 cm optics were measured to be 70\% through MS1 alone, 83\% through MS2 alone, and 63\% through the system of both optics.

As another measure of optical performance, the modulation transfer function (MTF) describes the image contrast as a function of frequency. For an eyepiece optic with the function of projecting a displayed image to infinity, it is more appropriate to characterize the performance in terms of angular resolution (in cycles/mrad) rather than spatial resolution. From the measured point spread functions, we calculate the MTF at various angles of incidence in Figures \ref{Fig:PSFs}e and \ref{Fig:PSFs}f. The experiment results (solid lines) show consistent performance over the entire field of view and good agreement with the simulated MTF (dashed lines). While the spatial resolution is similar for the 1 cm and 2 cm optics, the angular resolution of the 2 cm optics is much higher due to its longer effective focal length. For comparison, we plot the simulated MTF of the exemplary commercial refractive system (the same system depicted in Figure \ref{Fig:RayDiagrams}d) as darker dotted lines in Figure \ref{Fig:PSFs}f. While the refractive system has higher MTF at normal incidence, it drops rapidly at increasing angles of incidence, showing a true FoV of less than 20 degrees. This is in contrast to the designed meta-optics, which exhibit similar performance across the entire field of view. 

In general, the meta-optics experiment results exhibit excellent agreement with the ray tracing simulation. Upon close inspection, it is noted that under some conditions (namely $10^\circ$ through $30^\circ$ in the 1 cm optics) the experiment appears to slightly outperform the simulation. However, it should also be noted that the experimental performance is worse than the simulation at normal incidence. Therefore, we attribute the unexpectedly higher experimental performance to be due to slight misalignment which favors slightly off-axis angles. In the case of the 2 cm optics, the simulated results are consistently better than the experiment.

\begin{figure}[h!]
\centering\includegraphics[width=12cm]{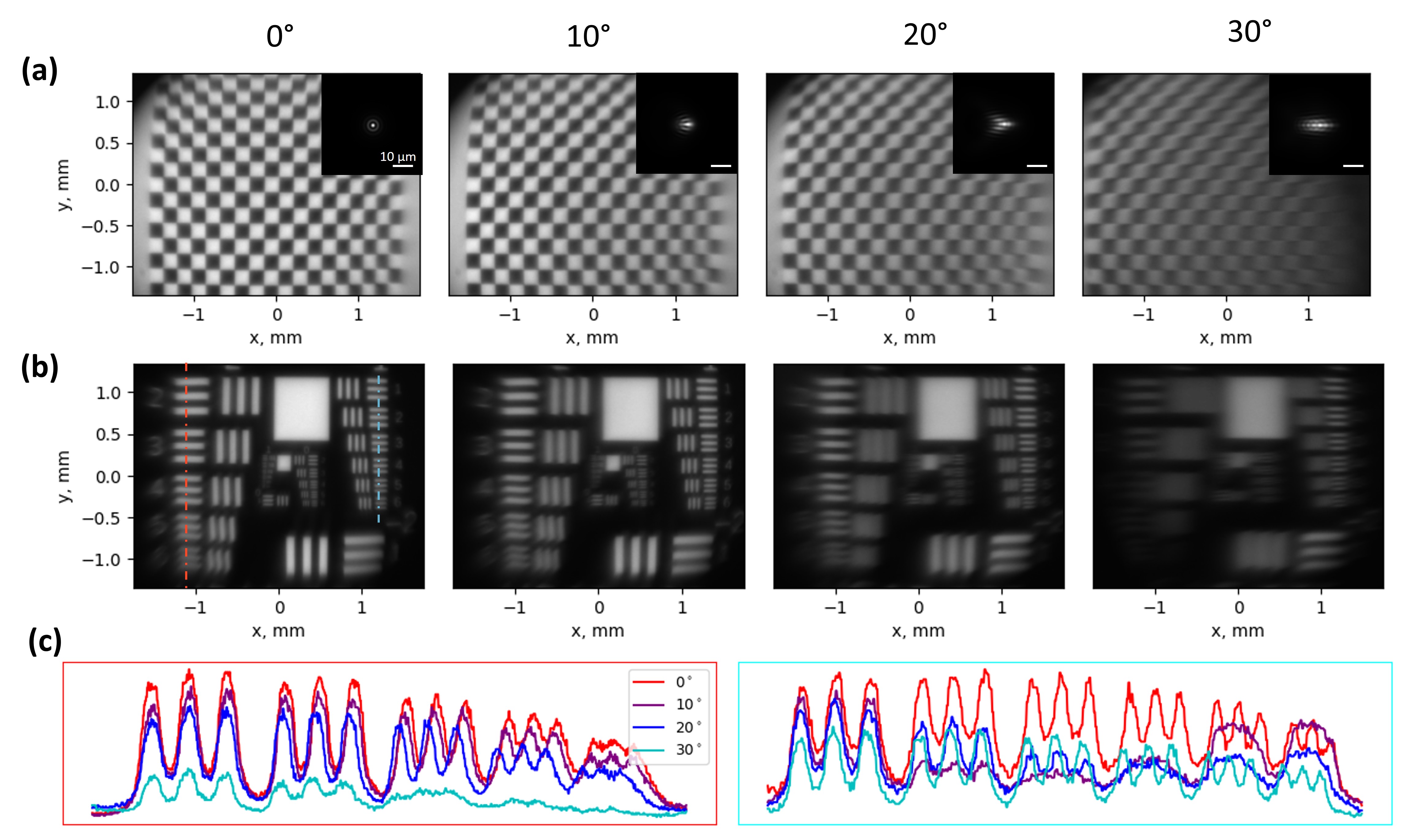}
\caption{Imaging results for the 1 cm aperture optics. (a) A checkerboard pattern was displayed on a micro-LED screen and imaged at angles from 0$^\circ$ to 30$^\circ$. The width of the re-imaged pattern is 3 cm. In the insets, we show the simulated PSF under for 632.8 nm illumination with 1 nm FWHM bandwidth. (b) The USAF resolution chart pattern was displayed on the micro-LED screen. The dashed red and cyan lines indicate where exemplary line cuts were taken. (c) Exemplary line cuts for each angle of incidence, illustrating the image contrast. }
\label{Fig:Imaging}
\end{figure}

In Figure \ref{Fig:Imaging}, we demonstrate the imaging quality of the 1 cm optics at wide field of view. We displayed the imaging pattern on a micro-LED display and re-imaged the pattern using a high NA objective (Nikon Plan Fluorite 20x) followed by a 633 nm narrowband filter (Thorlabs FLH632.8-1, FWHM ~1 nm). After passing through the objective, the size of the imaging object was approximately 3 mm wide. The displayed pattern was a checkerboard in Figure \ref{Fig:Imaging}(a) and a USAF resolution chart in Figure \ref{Fig:Imaging}(b). To assess performance across a wide field of view, the imaging object was rotated relative to the imaging system at the specified angle of incidence. In the insets in Figure \ref{Fig:Imaging}(c), we show exemplary line cuts through the red (left) and cyan (right) dashed lines to illustrate image contrast. Due to the slightly broad linewidth of the illumination source (FWHM 1 nm), the resolution is negatively affected by chromatic aberrations. In particular, off-axis performance is negatively affected by chromatic aberrations as the meta-optic phase gradient disperses illumination of different wavelengths. To elucidate this effect, we plot simulated polychromatic PSFs in the insets to Fig. \ref{Fig:Imaging}(a). These PSFs were simulated for 20 wavelengths between 630.5 nm and 635.0 nm, appropriately weighted to match the transmission spectrum of the narrowband filter. Lateral distortion of the PSF is apparent and worsens at increasing angle of incidence. Nonetheless, we demonstrate high-quality imaging up to 30$^\circ$ angle of incidence corresponding to 60$^\circ$ full field of view. 

\section{Discussion}

In this work, we balance trade-offs between compact form factor, aperture size, and field of view to achieve performance in a meta-optical doublet that is comparable to a commercial refractive eyepiece and meets common benchmarks of near eye displays. We note that the performance of the meta-optic doublet is not diffraction limited, but diffraction-limited performance is typically not required for AR/VR or night vision applications. In particular, human visual acuity of 20/20 corresponds to 1 arcmin angular resolution \cite{Xiong21}, or about 0.3 mrad. At this resolution, the contrast achieved with the meta-optic eyepiece is approximately 50\%, which is sufficient to resolve most features. As shown in Figure \ref{Fig:PSFs}f, the imaging quality is comparable to that of a similar commercial refractive lens at normal incidence and better at wide field of view. While the field of view demonstrated in the meta-optic eyepiece is wide relative to many camera lenses, extending the field of view further to a minimum of $100^\circ$ is desirable for a truly immersive AR/VR experience \cite{Chang20,Kress13}. 

The specifications of $80^\circ$ field of view for the 1 cm optics and $60^\circ$ field of view for the 2 cm optics were chosen considering the tradeoffs between form-factor, aperture size, and field of view. In this work, the primary limitation on field of view is the maximum diameter of meta-optics which can be reasonably fabricated. By increasing the diameter of MS1, the meta-optic can capture light at larger angles of incidence, which would increase the attainable field of view. As shown in several works, meta-optics achieving nearly $180^\circ$ field of view have been demonstrated, but only for small entrance aperture \cite{Mart20,Shal20}. In fact, it has been shown that in order to simultaneously achieve large aperture and diffraction-limited performance over wide field of view, it is necessary to increase the thickness of the optical system \cite{Li22,Miller23,Martins22}. Intuitively, this occurs because light of different angles of incidence must be spatially separated, which is easily achieved using a small entrance aperture to restrict light of different angles to interacting with different sections of the lens. In the case of large aperture, greater distance between the aperture and the lens is required to provide sufficient separation between the beams of different angles of incidence. 

In this work, the aperture of the meta-optics is practically limited by the feasible write time for electron beam lithography. Other fabrication techniques, including deep ultraviolet (DUV) lithography, provide much faster write time and are therefore more suitable for large apertures and mass production. However, the resolution achieved in our DUV facilities is around 250 nm, which is insufficient to fabricate the designed meta-atoms operating in visible with sub-wavelength periodicity. In the Supplementary Information, we detail efforts to overcome this issue by utilizing larger periodicity meta-atoms. We designed meta-atoms with 1100 nm periodicity and fabricated the full 2 cm aperture meta-optics using DUV lithography. However, due to aliasing issues introduced by insufficient phase sampling, the field of view of these meta-optics is limited to approximately $40^\circ$. Details of the DUV fabrication and experiment results are included in the Supplementary Information.

We note that the presented meta-optic doublet is designed for single-wavelength illumination at 633 nm, suitable primarily for monochromatic applications such as night vision. While much progress has been made in recent years to develop meta-optics with broadband operation in the visible regime \cite{Froch24,Tseng21,Huang22}, simultaneous achievement of broadband operation and wide field of view has yet to be demonstrated. For single-layer meta-optics, there are fundamental tradeoffs between device thickness, NA, and bandwidth \cite{Presutti20}. When extending this system to broadband operation, however, these limitations may be circumvented using a doublet configuration to provide additional degrees of freedom. In addition, high resolution over the entire field of view is not required due to the foveated nature of the human eye \cite{Curcio90, Saragadam24,Chang20}; that is, performance requirements at ultrawide field of view are relaxed, which may enable the extension of the design presented here to broadband applications. 

In conclusion, we demonstrated a large aperture, wide field of view meta-optic doublet eyepiece for near-eye display applications. Our design considers realistic constraints such as eye relief, pupil size, and display size. In incremental steps towards a large aperture meta-optic eyepiece, we designed a smaller system with 1 cm entrance aperture as a proof of concept as well as a full-scale 2 cm entrance aperture system. In both cases, the experimental performance of the system closely agrees with the design and exhibits consistent performance over at least $60^\circ$ full field of view. These findings represent promising results for the integration of meta-optics into full-scale near-eye display systems, including AR/VR and night vision.

\section{Materials and Methods}

\subsection{Metasurface Fabrication}
The meta-optics are fabricated in a silicon nitride layer on quartz substrate (500 $\mu$m thick). First, 750 nm SiN was deposited on the substrate via plasma-enhanced chemical vapor deposition (Oxford; Plasma Lab 100). The sample was coated with electron beam photoresist (ZEP-520A) and patterned using electron beam lithography (JEOL; JBX6300FS). Afterwards, we deposited 80 nm alumina (Al$_2$O$_3$) using an electron beam evaporator (CHA; SEC-600) and did liftoff with 1-methyl-2-pyrrolidionone to form a hard mask for etching.  The silicon nitride was then etched in an inductively coupled plasma etcher (Oxford; PlasmaLab 100, ICP-180) using fluorine-based gas chemistry. 

\subsection{Meta-Optics Characterization}

The PSF of the meta-optics was measured on a home-built setup. The illumination source was a HeNe laser (Newport N-LHP-131) at 632.8 nm wavelength with 1400 MHz linewidth (FWHM). The laser output was coupled to a single-mode fiber with approximately 300 nm diameter fiber core. A refractive lens was used to collimate the output. For measurements at various angles of incidence, the fiber output and collimator unit were mounted on a rotating arm to provide smooth rotation up to 40$^\circ$. An iris placed at the axis of rotation served as the pupil aperture. Each meta-optic was mounted on 3-axis translation stage for precise alignment. The focal spot was magnified via microscope objective (Nikon Plan Fluorite 20x, NA = 0.50, WD = 2.1 mm) followed by a tube lens. The output was measured on a GT1930C camera sensor with 5.86 $\mu$m per pixel resolution. The effective pixel resolution given the relay optics was calibrated by imaging an object of known size with the relay system. For the data shown in Fig. \ref{Fig:PSFs}a and \ref{Fig:PSFs}c, the exposure time was 83 $\mu$s, and for the data shown in Fig. \ref{Fig:PSFs}b and \ref{Fig:PSFs}d, the exposure time was 69 $\mu$s. The camera and relay optics were mounted on a two-axis stage for precise positioning of focal length and lateral translation to collect off-axis PSFs. Schematics and further details of the PSF characterization setup are available in the Supplementary Information.

The experimental MTFs, shown in Figs.\ref{Fig:PSFs}e and \ref{Fig:PSFs}f, were calculated from the corresponding normalized PSFs shown in Figs. \ref{Fig:PSFs}a and \ref{Fig:PSFs}b. First, to correct for nonzero background value, an estimate of the mean background value was obtained from a 200x200 pixel area near the corner of each image. This background value was then subtracted from the data. To calculate the MTF, we computed the 2D Fourier transform of the normalized and background-corrected PSF and took a line cut along the same axis as shown PSF cuts. The MTF was then converted from the spatial unit of lp/mm to the angular unit of cyc/mrad using the effective focal length, which is 5.84 mm for the 1 cm optics and 15.17 mm for the 2 cm optics.

For an imaging demonstration, we built an eye model consisting of a refractive lens of 2.5 mm focal length and the camera to collect the image, and both were mounted on the rotating arm. The meta-optics were mounted and aligned according to the design. For the imaging object, the desired pattern was displayed on a micro-LED screen and a narrowband filter (Thorlabs FL632.8-1, FWHM $\approx$1 nm) was used to filter illumination to the desired wavelength. The microscope objective was used to re-image the display object to a lateral size of approximately 3 mm. To measure across the entire FoV, the display, filter, and objective were translated laterally to cover the designed 8 mm display size corresponding to $40^\circ$ FoV. Further details of the experiment setup, including schematic diagrams and the meta-optic alignment procedure, are available in the Supplementary Information.

\medskip

\section*{Author Contributions}

T.G. designed the meta-optic doublet system and phase profiles in Zemax. F.Y., L.M., and H.Z. contributed to modeling and discussion. A.W.-S. and J.F. designed the meta-atom libraries and made the GDS files. J.F. fabricated the meta-optics. Q.T. and Z.H. assisted with developing the fabrication processes. A.W.-S., J.F., and Z.Z. conducted the experimental measurements. A.W.-S. and J.F. analyzed the experiment data. A.W.-S. and J.F. drafted the manuscript. For DUV Stepper lithography and etching, D.J. directed the fabrication process and B.S. ran all lithography optimizations, SEM's, and final product wafers. All authors contributed to technical discussion and review of the manuscript. A.M., J.H., and T.G. supervised the project and coordinated the research.

\section*{Acknowledgements}

We gratefully acknowledge assistance from the staff at UCSB Nanofabrication Facility for DUV Stepper lithography and etching. Biljana Stamenic ran all the lithography optimizations, SEM's, etch calibrations and final product wafers. Demis D. John  provided fabrication process direction, developed the initial process and updated process to use SiO-hardmasked Ru etching, and mask design reviews. William J. Mitchell developed the Ruthenium hard mask process, and provided recipes and process input on how to deposit and etch both the the Ru and SiO$_2$ using his established processes. The lithography masks were purchased from Digidat, Inc. 

We gratefully acknowledge fabrication assistance from several individuals at the University of Washington. Erik Petersen deposited SiN and prepared wafers for future fabrication steps, and Arnab Manna for diced the completed wafers after DUV stepper fabrication. In addition, we acknowledge Tina Teichmann for assisting with the experimental measurements.   

\section*{Funding}
\noindent Funding for this work was supported by the DARPA-ENVision program.

\noindent Part of this work was conducted at the Washington Nanofabrication Facility / Molecular Analysis Facility, a National Nanotechnology Coordinated Infrastructure (NNCI) site at the University of Washington with partial support from the National Science Foundation via awards NNCI-1542101 and NNCI-2025489. 
Part of this work was performed in the UCSB Nanofabrication Facility, an open access laboratory.

\section*{Disclosures}
\noindent A.M. is a co-founder of Tunoptix, which is commercializing similar meta-optics in the visible. T.G. and J.H. are co-founders of 2Pi Inc., a company commercializing metasurface optics.

\section*{Data availability} 
\noindent Data underlying the results presented in this paper are not publicly available at this time but may be obtained from the authors upon reasonable request.

\bigskip



\printbibliography[title={References}]  

\end{document}